\documentclass[12pt]{article}

\usepackage{graphicx}
\begin{document}

\begin{center}
{\bf Einstein$-$Gauss$-$Bonnet gravity with rational nonlinear electrodynamics} \\
\vspace{5mm} S. I. Kruglov
\footnote{E-mail: serguei.krouglov@utoronto.ca}
\underline{}
\vspace{3mm}

\textit{Department of Physics, University of Toronto, \\60 St. Georges St.,
Toronto, ON M5S 1A7, Canada\\
Department of Chemical and Physical Sciences, University of Toronto,\\
3359 Mississauga Road North, Mississauga, ON L5L 1C6, Canada} \\
\vspace{5mm}
\end{center}
\begin{abstract}
We obtain an exact spherically symmetric and magnetically charged black hole solution in 4D Einstein$-$Gauss$-$Bonnet gravity coupled with rational nonlinear electrodynamics. The thermodynamics of our model is studied. We calculate the Hawking temperature and the heat capacity of the black hole. The phase transitions occur in the point where the Hawking temperature possesses an extremum. We show that black holes are thermodynamically stable at some range of event horizon radii when the heat capacity is positive. The logarithmic correction to the Bekenstein$-$Hawking entropy is obtained.
\end{abstract}


The string theories give rise to effective models of gravity in higher dimensions with higher order curvature terms in the action when approaching the low energy limit. The Gauss$-$Bonnet (GB) Lagrangian, including higher order curvature terms, in four dimensions ($4D$), can be represented as a total derivative and, therefore, it possesses a topological nature. It was shown that the GB coupling constant $\alpha$ can be re-scaled as $\alpha/(D-4)$. Then in the limit $D \rightarrow 4$ there will be a contribution to the theory in $4D$ without singularities \cite{Glavan} and the Einstein$-$Gauss$-$Bonnet (EGB) theory leads to non-trivial dynamics. It should be noted that the 4D EGB gravity theory was debated in \cite{Tekin}-\cite{Hohmann}. But the dimensional regularization of \cite{Glavan} is justified for some class of spherically-symmetric metric \cite{Aoki}, \cite{Aoki1}.
The EGB theory of gravity has the attractive feature such as a repulsive gravitational force at short distances, and the GB term can be considered as a quantum correction to gravity \cite{Deser}, \cite{Wheeler}. The singularity problem in BH physics is absent because the gravitational force is repulsive at small distances. It is worth noting that 4D EGB theory preserves the number of graviton degrees of freedom, and as a result, it is free from the Ostrogradsky instability. The $4D$ EGB gravity theory recently received much attention (an incomplete list) \cite{Fernandes}- \cite{Wei1}. In this paper we find a black hole solution in the $4D$ EGB model coupled to rational nonlinear electrodynamics (RNED) as proposed in \cite{Kruglov1}. It is worth noting that nonlinear electrodynamics (NED) plays very important role in cosmology and black hole physics \cite{Bronnikov}-\cite{Kruglov}.



The action of the EGB gravity in D-dimensions coupled to RNED is given by
\begin{equation}
I=\int d^Dx\sqrt{-g}\left[\frac{1}{16\pi G}\left(R+ \frac{\alpha}{D-4}{\cal L}_{GB}\right)+{\cal L}_{RNED}\right],
\label{1}
\end{equation}
where $\alpha$ possesses the dimension of (length)$^2$ and the Lagrangian of RNED, proposed in \cite{Kruglov1}, is
\begin{equation}
{\cal L}_{RNED} = -\frac{{\cal F}}{1+2\beta{\cal F}},
 \label{2}
\end{equation}
with the parameter $\beta$ ($\beta\geq 0$) having the dimension of (length)$^4$, ${\cal F}=(1/4)F_{\mu\nu}F^{\mu\nu}=(B^2-E^2)/2$, $F_{\mu\nu}=\partial_\mu A_\nu-\partial_\nu A_\mu$ is the field strength tensor.
The GB Lagrangian reads
\begin{equation}
{\cal L}_{GB}=R^{\mu\nu\alpha\beta}R_{\mu\nu\alpha\beta}-4R^{\mu\nu}R_{\mu\nu}+R^2.
\label{3}
\end{equation}
The variation of action (1) with respect to the metric results in field equations
\begin{equation}
R_{\mu\nu}-\frac{1}{2}g_{\mu\nu}R+\frac{\alpha}{D-4}H_{\mu\nu}=-8\pi GT_{\mu\nu},
\label{4}
\end{equation}
where
\begin{equation}
H_{\mu\nu}=2\left(RR_{\mu\nu}-2R_{\mu\alpha}R^\alpha_{~\nu}-2R_{\mu\alpha\nu\beta}R^{\alpha\beta}-
R_{\mu\alpha\beta\gamma}R^{\alpha\beta\gamma}_{~~~\nu}\right)-\frac{1}{2}{\cal L}_{GB}g_{\mu\nu}.
\label{5}
\end{equation}
The symmetrical energy-momentum tensor of RNED is given by
\begin{equation}
T_{\mu\nu}=-\frac{F_\mu^{~\alpha}F_{\nu\alpha}}{(1+2\beta{\cal F})^{2}}
-g_{\mu\nu}{\cal L}_{RNED}.
\label{6}
\end{equation}
In the following we consider a magnetic BH with the spherically symmetric field. The static and spherically symmetric metric is given by
\begin{equation}
ds^2=-f(r)dt^2+\frac{1}{f(r)}dr^2+r^2(d\vartheta^2+\sin^2\vartheta d\phi^2).
\label{7}
\end{equation}
Taking into consideration that the electric charge $q_e=0$, ${\cal F}=q_m^2/(2r^4)$ ($q_m$ is a magnetic charge), one obtains from Eq. (6) the magnetic energy density \cite{Kruglov1}
\begin{equation}
\rho=T_t^{~t}=\frac{B^2}{2(\beta B^2+1)}=\frac{q_m^2}{2(r^4+\beta q_m^2)}.
\label{8}
\end{equation}
The symmetrical energy-momentum tensor (6) with the spherical symmetry leads to $T_t^{~t}=T_r^{~r}$. Then we obtain the radial pressure $p_r=-T_r^{~r}=-\rho$. The tangential pressure is defined by $p_\perp=-T_\vartheta^{~\vartheta}=-T_\phi^{~\phi}$ and is \cite{Dymnikova}
\begin{equation}
p_\perp=-\rho-\frac{r}{2}\rho'(r),
\label{9}
\end{equation}
where the prime denotes the derivative with respect to the argument. The Weak Energy Condition (WEC) is satisfied when $\rho\geq 0$ and $\rho+p_k\geq 0$ (k=1,2,3) \cite{Hawking}, i.e. the energy density is positive. In accordance with Eq. (8) $\rho\geq 0$. From Eq. (8) one finds
\begin{equation}
\rho'(r)=-\frac{2q_m^2r^3}{(r^4+\beta q_m^2)^2}\leq 0.
\label{10}
\end{equation}
As a result WEC $\rho\geq 0$, $\rho+p_r\geq 0$, $\rho+p_\perp\geq 0$ is satisfied for the model considered. The Dominant Energy Condition (DEC) holds if and only if \cite{Hawking} $\rho\geq 0$, $\rho+p_k\geq 0$, $\rho-p_k\geq 0$ which includes WEC. Therefore, we need only to verify the condition $\rho-p_\perp\geq 0$. From Eqs. (8)-(10) we obtain
\begin{equation}
\rho-p_\perp=\frac{\beta q_m^4}{(r^4+\beta q_m^2)^2}\geq 0.
\label{11}
\end{equation}
Thus, DEC is satisfied and, as a result, the sound speed is less than the speed of light.
The Strong Energy Condition (SEC) is satisfied if $\rho+\sum_{k=1}^3 p_k\geq 0$ \cite{Hawking}. Making use of Eqs. (8)-(10) one finds
\begin{equation}
\rho+\sum_{k=1}^3 p_k=\rho+3p=\frac{q_m^2(r^4-\beta q_m^2)}{(r^4+\beta q_m^2)^2}.
\label{12}
\end{equation}
It follows from Eq. (12) that SEC is not fulfilled.

We consider the limit $D \rightarrow 4$ and at $\mu=\nu=t$ the field equation (4) gives
\begin{equation}
r(2\alpha f(r)-r^2-2\alpha)f'(r)-(r^2+\alpha f(r)-2\alpha)f(r)+r^2-\alpha=\frac{q_m^2r^4G}{r^4+\beta q_m^2}.
\label{13}
\end{equation}
The solution to Eq. (13) is given by
\begin{equation}
f(r)=1+\frac{r^2}{2\alpha}\left(1-\sqrt{1+\frac{8M\alpha G}{r^3}+\frac{\alpha q_m^{3/2}h(r)G}{\sqrt{2}\beta^{1/4}r^3}}\right),
\label{14}
\end{equation}
\[
h(r)=\ln\frac{r^2-\sqrt{2}\sqrt[4]{\beta q_m^2}r+\sqrt{\beta q_m^2}}{r^2+\sqrt{2}\sqrt[4]{\beta q_m^2}r+\sqrt{\beta q_m^2}}+2\arctan\left(1+\frac{\sqrt{2}r}{\sqrt[4]{\beta q_m^2}}\right)
\]
\begin{equation}
-2\arctan\left(1-\frac{\sqrt{2}r}{\sqrt[4]{\beta q_m^2}}\right).
\label{15}
\end{equation}
For convenience we introduce the unitless variable $x=r/\sqrt[4]{\beta q_m^2}$. Then solution (14) becomes
\begin{equation}
f(x)=1+cx^2-c\sqrt{x^4+x(a+bg(x))},
\label{16}
\end{equation}
where
\[
a=\frac{8M\alpha G}{\beta^{3/4}q_m^{3/2}},~~~b=\frac{\alpha G}{\sqrt{2}\beta},~~~c=\frac{\sqrt{\beta}q_m}{2\alpha},
\]
\begin{equation}
g(x)=\ln\frac{x^2-\sqrt{2}x+1}{x^2+\sqrt{2}x+1}+2\arctan(1+\sqrt{2}x)-2\arctan(1-\sqrt{2}x).
\label{17}
\end{equation}
The constant of integration was chosen to be $8M\alpha G$ ($M$ is the Schwarzschild BH mass) to have the charged BH solution of GR. In addition, we use the sign minus before the square root in Eqs. (14) and (16). In this case the BH is stable and without ghosts \cite{Deser}.
The asymptotic of the metric function $f(r)$ (14) is given by
\begin{equation}
f(r)=1+\sqrt{\frac{2MG}{\alpha}}\sqrt{r}+\frac{r^2}{2\alpha}+{\cal O}(r^3)~~~~r\rightarrow 0,
\label{18}
\end{equation}
\begin{equation}
f(r)=1-\frac{2mG}{r}+\frac{Gq_m^2}{r^2}+{\cal O}(r^{-3})~~~~r\rightarrow \infty,~~~~
m=M+\frac{\pi q_m^{3/2}}{4\sqrt{2}\beta^{1/4}}\equiv M+m_m,
\label{19}
\end{equation}
where $m$ is the total mass of the BH which includes the Schwarzschild mass $M$ and the electromagnetic mass $m_m=\int_0^\infty r^2\rho dr=\pi q_m^{3/2}/(4\sqrt{2}\beta^{1/4})$. Equation (18) shows that we have the regular BH because $f(r)\rightarrow 1$ as $r\rightarrow 0$. According to Eq. (19) the Reissner$-$Nordstr\"{o}m behavior of the charged BH holds at infinity.

The limit $\beta\rightarrow 0$, to have EGB coupled to Maxwell electrodynamics, should be made in Eq. (13) (before the integration). Then the solution to Eq. (13) at $\beta=0$ is given by \cite{Fernandes}
\begin{equation}
f(r)=1+\frac{r^2}{2\alpha}\left(1-\sqrt{1+\frac{8M\alpha G}{r^3}-\frac{4\alpha q_m^2G}{r^4}}\right).
\label{20}
\end{equation}
It is worth mentioning that the limit $r\rightarrow 0$ in Eq. (20) leads to a non-physical complex value of the metric function $f(r)$, but the limit $r\rightarrow 0$ in Eq. (14) (and (16)) leads to the reasonable value $f(0)=1$.
The plot of the function (16) is depicted in Fig. 1.
\begin{figure}[h]
\includegraphics[height=4.0in,width=4.0in]{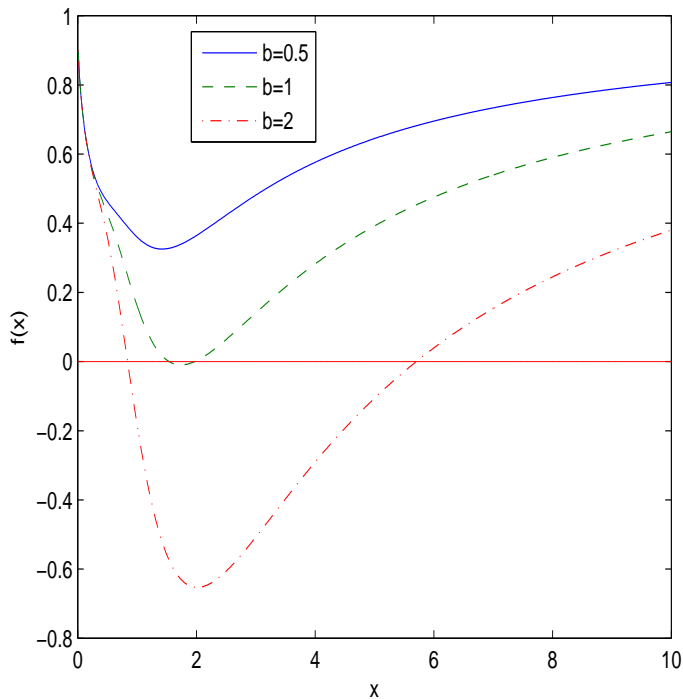}
\caption{\label{fig.1}The plot of the function $f(x)$ for $a=c=1$.}
\end{figure}
Figure 1 shows that there can be two horizons, one (the extreme) horizon or no horizons (the BH does not exist).


Consider the BH thermodynamics and the thermal stability of the BH. The Hawking temperature is given by
\begin{equation}
T_H(r_+)=\frac{f'(r_+)}{4\pi},
\label{21}
\end{equation}
where $r_+$ is the event horizon radius defined by the root of the equation $f(r_+)=0$. With the help of Eq. (16), with the variable $x=r/\sqrt[4]{\beta q_m^2}$, one finds the Hawking temperature
\begin{equation}
T_H(x_+)=\frac{1}{4\pi \sqrt[4]{\beta q_m^2}}\left(\frac{(2cx_+^2-1)(1+x_+^4)-4\sqrt{2}bc^2x_+^4}{2x_+(1+x_+^4)(1+cx_+^2)}\right),
\label{22}
\end{equation}
where we replaced parameter $a$ from equation $f(x_+)=0$.
The plot of the dimensionless function $T_H(x_+)\sqrt[4]{\beta q_m^2}$ is depicted in Fig. 2.
\begin{figure}[h]
\includegraphics[height=4.0in,width=4.0in]{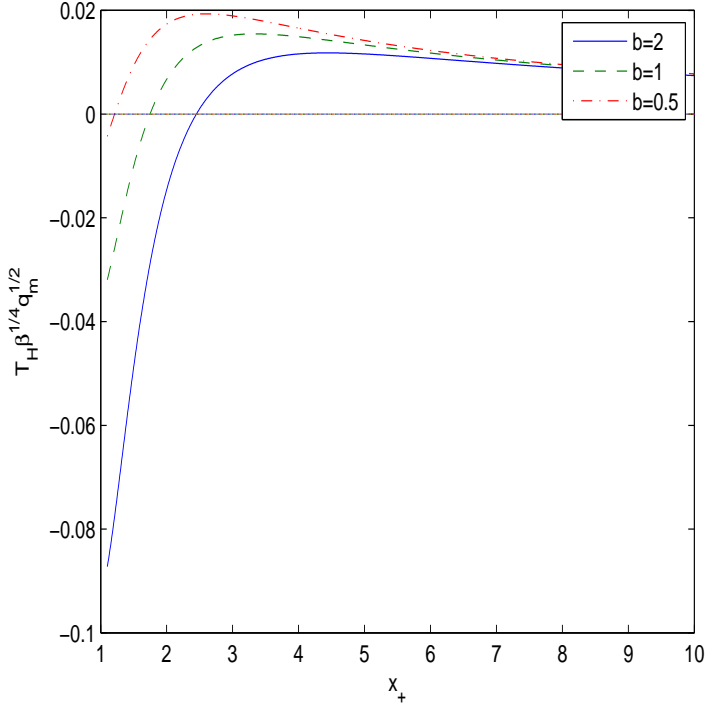}
\caption{\label{fig.2}The plot of the function $T_H(x_+)\sqrt[4]{\beta q_m^2}$ at $c=1$.}
\end{figure}
Figure 2 shows that the Hawking temperature is positive for some range of $x_+$ while the BH does not exist when the Hawking temperature is negative.  Solving equation $f(x_+)=0$ we obtain the BH gravitational
mass
\begin{equation}
M(x_+)=\frac{\beta^{3/4}q_m^{3/2}}{8\alpha G}\left(\frac{1+2cx_+^2}{c^2x_+}-bg(x_+)\right).
 \label{23}
\end{equation}
Making use of first law of BH thermodynamics
\begin{equation}
dM(x_+)=T_H(x_+)dS+\phi dq,
\label{24}
\end{equation}
one can find the entropy at the constant charge
\begin{equation}
S=\int \frac{dM(x_+)}{T_H(x_+)}=\int \frac{1}{T_H(x_+)}\frac{\partial M(x_+)}{\partial x_+}dx_+.
\label{25}
\end{equation}
The heat capacity allows us to study the stability of the BH. Making use of Eq. (25) we obtain the expression for the heat capacity
\begin{equation}
C_q(x_+)=T_H\left(\frac{\partial S}{\partial T_H}\right)_q=\frac{\partial M(x_+)}{\partial T_H(x_+)}=\frac{\partial M(x_+)/\partial x_+}{\partial T_H(x_+)/\partial x_+}.
\label{26}
\end{equation}
With the help of Eqs. (22) and (23) one finds
\begin{equation}
\frac{\partial M(x_+)}{\partial x_+}=\frac{\beta^{3/4}q_m^{3/2}}{8\alpha G}\left(\frac{2cx_+^2-1}{c^2x_+^2}-\frac{4\sqrt{2}bx_+^2}{1+x_+^4}\right).
\label{27}
\end{equation}
\begin{equation}
\frac{\partial T_H(x_+)}{\partial x_+}=\frac{1}{4\pi \sqrt[4]{\beta q_m^2}}\biggl(\frac{5cx_+^2-2c^2x_+^4+1}{2x_+^2(1+cx_+^2)^2}
-\frac{2\sqrt{2}bc^2(cx_+^4-3cx_+^8-x_+^6+3x_+^2}{(1+x_+^4)^2(1+cx_+^2)^2}\biggr).
\label{28}
\end{equation}
In accordance with Eq. (26) the heat capacity has a singularity when the Hawking temperature possesses an extremum ($\partial T_H(x_+)/\partial x_+=0$).
Equations (22) and (26) show that at some point, $x_+=x_1$, the Hawking temperature and heat capacity become zero and a first-order phase transition takes place. At this point $x_1$ the BH remnant occurs, where the BH mass is not zero but the Hawking temperature and the heat capacity vanish. At the event horizon radius $x_+=x_2$ where $\partial T_H(x_+)/\partial x_+=0$ the heat capacity has the discontinuity. Therefore, in the range $x_2>x_+>x_1$ BHs are locally stable. At $x_+>x_2$ the BH evaporates and becomes unstable.

Making use of Eqs. (26), (27) and (28) we plotted the heat capacity versus the variable $x_+$ in Fig. 3.
\begin{figure}[h]
\includegraphics[height=4.0in,width=4.0in]{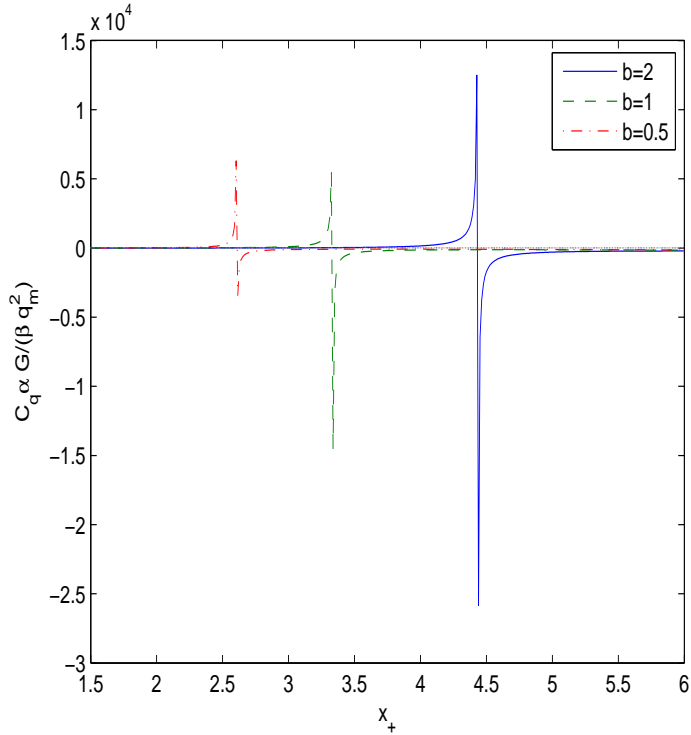}
\caption{\label{fig.3}The plot of the function $C_q(x_+)\alpha G/(\beta q_m^2)$ at $c=1$. }
\end{figure}
Figure 3 shows that the BH is stable in the interval $x_2>x_+>x_1$ where the heat capacity is positive.
The singularity in the heat capacity occurs at the points where the Hawking temperature has the extremum and the second-order phase transition takes place.

From Eqs. (22), (25) and (27)) we obtain the entropy
\begin{equation}
S=\frac{4\pi \alpha}{G}\int\frac{1+cx_+^2}{x_+}dx_+=
\frac{\pi r_+^2}{G}+\frac{4\pi\alpha}{G}\ln\left(\frac{r_+}{\sqrt[4]{\beta q_m^2}}\right).
\label{29}
\end{equation}
It follows from Eq. (29) that we have the logarithmic correction to the Bekenstein$-$Hawking entropy in our model of 4D EGB gravity coupled to RNED. It is worth  noting that a logarithmic term was obtained in GR with a conformal
anomaly \cite{Cai}, \cite{Cai1}. Therefore, we can treat the logarithmic correction as mimicking the quantum corrections.
According to Eq. (29), at $\alpha=0$ one comes to the Bekenstein$-$Hawking entropy.
For the large event horizon radius $x_+$ (for massive BHs) the logarithmic correction to the area low is not important but for light BHs (for small $x_+$) one can not ignore the logarithmic term in the entropy.

Equation (29) shows that at some values of $\alpha$, $\beta$, $q_m$  and $r_+$ the entropy vanishes. The corresponding event horizon radius is the solution of equation ($S=0$) $r_+^4\exp(r_+^2/\alpha)=\beta q_m^2$, which is $r_+=\sqrt{2\alpha W_0(\sqrt{\beta} q_m/(2\alpha))}$, where $W_0(x)$ is the Lambert function. At $r_+<\sqrt{2\alpha W_0(\sqrt{\beta} q_m/(2\alpha)}$ the entropy is negative. The negative entropy of BHs was discussed in \cite{Odintsov}.


We summarize the results obtained as follows. The exact spherically symmetric and magnetically charged BH solution in 4D EGB gravity coupled with RNED was obtained. It was demonstrated that WEC and DEC are satisfied. The thermodynamics and the thermal stability of magnetized BHs were investigated. The Hawking temperature and the heat capacity of BHs were calculated  in the EGB model coupled to RNED. The phase transitions occur in the points where the Hawking temperature possesses the extremum. We have shown that BHs are thermodynamically stable at some range of event horizon radii when the heat capacity and the Hawking temperature are positive. The heat capacity diverges at some event horizon radii where the phase transitions of the second-order occur. The logarithmic correction to the Bekenstein$-$Hawking entropy was established.

\end{document}